\documentclass[aip,floatfix,jcp,reprint,longbibliography]{revtex4-1}
\usepackage[latin9]{inputenc}
\usepackage{amsmath}
\usepackage{amssymb}
\usepackage[unicode=true,
 bookmarks=false,
 breaklinks=false,pdfborder={0 0 1},backref=section,colorlinks=false]
 {hyperref}

\makeatletter

\@ifundefined{textcolor}{}
{%
  \definecolor{BLACK}{gray}{0}
  \definecolor{WHITE}{gray}{1}
  \definecolor{RED}{rgb}{1,0,0}
  \definecolor{GREEN}{rgb}{0,1,0}
  \definecolor{BLUE}{rgb}{0,0,1}
  \definecolor{CYAN}{cmyk}{1,0,0,0}
  \definecolor{MAGENTA}{cmyk}{0,1,0,0}
  \definecolor{YELLOW}{cmyk}{0,0,1,0}
}

\usepackage{graphicx}          
\usepackage{amsmath}    
\usepackage{amssymb}    
\usepackage{hyperref}   
\usepackage[capitalise]{cleveref}   

\usepackage{enumitem}
\usepackage{amsthm}
\usepackage{float}
\usepackage[capitalise]{cleveref}
\usepackage{appendix}
\usepackage{color}
\usepackage{soul}
\usepackage{bm}

\makeatother

\begin{document}
\title{Special Topic: Markov Models of Molecular Kinetics}
\author{Frank Noé}
\email{frank.noe@fu-berlin.de}

\affiliation{Department of Mathematics and Computer Science, Freie Universität
Berlin, Berlin, Germany}
\affiliation{Department of Physics, Freie Universität Berlin, Berlin, Germany}
\author{Edina Rosta}
\email{edina.rosta@kcl.ac.uk}

\affiliation{Department of Chemistry, Kings College London, London, England}
\maketitle

\paragraph{Introduction}

The \emph{Journal of Chemical Physics} article collection on \href{https://aip.scitation.org/toc/jcp/collection/10.1063/jcp.2019.MMMK.issue-1}{Markov Models of Molecular Kinetics (MMMK)}
features recent advances developing and using Markov State Models
(MSMs) \citep{SchuetteFischerHuisingaDeuflhard_JCompPhys151_146,SwopePiteraSuits_JPCB108_6571,NoeHorenkeSchutteSmith_JCP07_Metastability,ChoderaEtAl_JCP07,BucheteHummer_JPCB08,PrinzEtAl_JCP10_MSM1}
in atomistic molecular simulations and related applications -- see
\citep{BowmanPandeNoe_MSMBook,SarichSchuette_MSMBook13,ChoderaNoe_COSB14_MSMs,HusicPande_JACS18_MSMReview}
for recent MSM reviews. MSMs have been an important driving force
in molecular dynamics (MD), as they facilitate divide-and-conquer
integration of short, distributed MD simulations into long-timescale
predictions, they are conceptually simple and provide readily-interpretable
models of kinetics and thermodynamics.

Most MSM estimation approaches proceed by a sequence, or pipeline,
of data processing steps that is also represented by MSM software
packages \citep{SchererEtAl_JCTC15_EMMA2,DoerrEtAl_JCTC16_HTMD,HarriganEtAl_BJ17_MSMbuilder},
and typically includes:
\begin{enumerate}
\item \textbf{Featurization}: The MD coordinates are transformed into features,
such as residue distances, contact maps or torsion angles \citep{HumphreyDalkeSchulten_JMG96_VMD,McGibbon_BJ15_MDTraj,SchererEtAl_JCTC15_EMMA2,DoerrEtAl_JCTC16_HTMD},
that form the input of the MSM analysis.
\item \textbf{Dimension reduction}: The dimension is reduced to much fewer
(typically 2-100) slow collective variables (CVs), \citep{NoeNueske_MMS13_VariationalApproach,NueskeEtAl_JCTC14_Variational,PerezEtAl_JCP13_TICA,SchwantesPande_JCTC13_TICA,Molgedey_94,ZieheMueller_ICANN98_TDSEP,Mezic_NonlinDyn05_Koopman,SchmidSesterhenn_APS08_DMD,WilliamsKevrekidisRowley_JNS15_EDMD,TuEtAl_JCD14_ExactDMD,NoeClementi_COSB17_SlowCVs}.
The resulting coordinates may be scaled, in order to embed them in
a metric space whose distances correspond to some form of dynamical
distance \citep{NoeClementi_JCTC15_KineticMap,NoeClementi_JCTC16_KineticMap2}.
\item \textbf{Discretization}: The space may be discretized by clustering
the projected data \citep{BowmanPandeNoe_MSMBook,SchererEtAl_JCTC15_EMMA2,HusicPande_JCTC17_Ward,SheongEtAl_JCTC15_APM,ChoderaEtAl_JCP07,WuNoe_JCP15_GMTM,HarriganPande_bioRxiv17_LandmarkTICA,WeberFackeldeySchuette_JCP17_SetfreeMSM},
typically resulting in 100-1000 discrete ``microstates''.
\item \textbf{MSM estimation}: A transition matrix or rate matrix describing
the transition probabilities or rate between the discrete states at
some lag time $\tau$ is estimated \citep{Bowman_JCP09_Villin,PrinzEtAl_JCP10_MSM1,BucheteHummer_JPCB08,TrendelkampSchroerEtAl_InPrep_revMSM}.
\item \textbf{Coarse-graining}: In order to get an easier interpretable
kinetic model, the MSM from step 5 is often coarse-grained to a few
states \citep{DeuflhardWeber_LAA05_PCCA+,KubeWeber_JCP07_CoarseGraining,YaoHuang_JCP13_Nystrom,FackeldeyWeber_WIAS17_GenPCCA,GerberHorenko_PNAS17_Categorial,HummerSzabo_JPCB15_CoarseGraining,OrioliFaccioli_JCP16_CoarseMSM,NoeEtAl_PMMHMM_JCP13,Schuette_HandbookNumAnal_2003}.
\end{enumerate}
Some method skip or combine some of these steps, novel machine learning
methods attempt to integrate most or all of them in an end-to-end
learning framework.

Key in much of the methodological progress in Markov modeling has
been the mathematical theory of conformation dynamics pioneered by
Schütte \citep{SchuetteFischerHuisingaDeuflhard_JCompPhys151_146}
and further developed by many contributors. This theory models the
dynamics of molecules by a Markov propagator $\mathcal{T}(\tau)$
that describes how an ensemble of molecules $\rho_{0}$ evolves in
a time step $\tau$. 
\[
\rho_{\tau}=\mathcal{T}(\tau)\rho_{0}.
\]
The MSM transition matrix is a discrete version of $\mathcal{T}(\tau)$.
Discretization in high-dimensional spaces is difficult to impossible,
so it is important to understand the structure underlying these dynamics
in order to make the MSM estimation problem feasible. If the dynamics
are in equilibrium, this propagation can further be approximated by
a sum of processes $\psi_{i}$ that relax the initial distribution
towards the equilibrium distribution with characteristic time scales
$t_{i}$.
\begin{equation}
\rho_{\tau}(\mathbf{x})\approx\sum_{i}\mathrm{e}^{-\tau/t_{i}}\langle\psi_{i},\rho_{0}\rangle\psi_{i}(\mathbf{x})\label{eq:EVD}
\end{equation}
As $\mathrm{e}^{-\tau/t_{i}}$ decays exponentially fast in the time
step $\tau$, only few terms are needed for Eq. (\ref{eq:EVD}) to
be an accurate description if we focus on the long-time dynamics,
i.e., the kinetics. The key insight from this theory is no less that
Markov modeling is possible even for complicated and very high-dimensional
molecular systems: We cannot sample or discretize truly high-dimensional
spaces, but we can do that for metastable molecular systems because
we are ultimately only interested in the low-dimensional manifold
spanned by a few eigenfunctions $\psi_{i}$ of the Markov operator.
Characterizing this manifold more compactly than by modeling all relevant
eigenfunctions $\psi_{i}$ explicitly is subject of current research
\citep{BittracherEtAl_arXiv2017_TransitionManifols}.

An important cornerstone for improving MSM estimators, developing
new ones and for turning MSM estimation into generic machine learning
problem that can be combined with kernel machines or neural networks,
is the development of variational optimization methods. The variational
approach for conformation dynamics (VAC) \citep{NoeNueske_MMS13_VariationalApproach,NueskeEtAl_JCTC14_Variational}
shows that eigenvalues of MSMs (the same is true for other linear
Markovian models such as TICA) systematically underestimates timescales
$t_{i}$ and eigenvalues $\mathrm{e}^{-\tau/t_{i}}$, and defines
a variational score -- essentially the sum of eigenvalues of an estimated
Markov model -- that ought to be maximized to optimally approximate
the unknown eigenvalues and eigenfunctions in (\ref{eq:EVD}). The
variational formulation is key to many contributions in the MMMK collection,
and remains to be the subject of further development and application,
e.g. in the context of MSM hyperparameter optimization \citep{McGibbonPande_JCP15_CrossValidation,HusicPande_JCTC17_Ward}.

While VAC only describes the scenario of equilibrium dynamics, i.e.
where our dynamics have a unique equilibrium distribution and obey
detailed balance, much recent research has focused on non-equilibrium
Markov models \citep{KoltaiEtAl_Computation18_NonrevMSM,ReuterEtAl_JCPMSM19_Schur,KnochSpeck_JCPMSM19_Noneq,KnochSpeck_NJP15_NoneqMSM,WuNoe_VAMP,MardtEtAl_VAMPnets}.
While nonequilibrium MSM studies are still in their infancy, several
theoretical principles are known in oder to make progress here. An
important framework for the description of such processes is that
of nonequilibrium work as covered by the Jarzynski fluctuation theorem
\citep{Jarzynski_PRL97}. From a machine learning and optimization
perspective, we can replace the eigenvalue decomposition in (\ref{eq:EVD})
with a singular value decomposition of the operator whose components
can be approximated with the variational approach of Markov processes
(VAMP) \citep{WuNoe_VAMP}. VAMP is exploited in several contributions
in the MMMK collection.

\paragraph{Feature selection}

One step in MSM estimation that had not yet undergone systematic analysis
or optimization is the selection of features used as an input. For
solvated molecules, roto-translationally invariant features were usually
chosen based on what works best for a given application -- including
intramolecular distances, angles, contact matrices or features implicitly
defined by pairwise metrics, such as minimal root mean square distance.
The VAC approach has previously been invoked to define variationally
optimal features for short peptides, in the spirit of defining optimized
basis sets in quantum chemistry \citep{VitaliniNoeKeller_JCTC15_BasisSet}.
In the MMMK collection, Scherer et al. \citep{SchererEtAl_JCPMSM19_VAMPFeatures}
propose to use the VAMP approach to variationally score different
candidates of features for a given MD analysis task. Considering a
large list of candidate features and all 12 fast-folding protein simulations
published by DESRES \citep{LindorffLarsenEtAl_Science11_AntonFolding},
the authors of \citep{SchererEtAl_JCPMSM19_VAMPFeatures} conclude
that a combination of residue-residue contact signals that decay exponentially
in the distance and backbone torsions performs best for protein folding.

\paragraph{Slow collective variables}

A major leap forward in MSM construction was the finding that machine-learning
methods that identify a manifold of slow collective variables (CVs)
\citep{NoeClementi_COSB17_SlowCVs}, such as the time-lagged independent
analysis (TICA) method \citep{Molgedey_94}, led to superior MSMs
\citep{PerezEtAl_JCP13_TICA,SchwantesPande_JCTC13_TICA,NaritomiFuchigami_JCP11_TICA}.
Intuitively, this success is due to the fact that MSMs aim a modeling
the kinetics between metastable states, and first reducing the dimension
to the manifold of slow (kinetic) processes makes subsequent geometric
operations such as clustering much simpler and faster than directly
working in a high-dimensional space. Theoretically, these methods
can indeed be derived from VAC \citep{NoeNueske_MMS13_VariationalApproach,PerezEtAl_JCP13_TICA,NueskeEtAl_JCTC14_Variational},
and thus be showed to variationally approximate the eigenfunctions
of the Markov propagator (\ref{eq:EVD}).

Several papers in the MMMK collection develop this approach further.
Karasawa et al. \citep{KarasawaEtAl_JCPMSM19_RMA} propose and extension
to relaxation mode analysis (RMA) \citep{TakanoMiyashita_JPSJ95_RelaxationModeAnalysis},
a close sibling of TICA, in which one first solves an eigenvalue problem
of the time correlation matrix of features to identify the manifold
of slow CVs, and then finds the subspace in which the matrix is positive
definite, promoting numerically stable estimate of relaxation rates.

A close relative of VAC is the spectral gap optimization of order
parameters (SGOOP) developed by Tiwary \citep{TiwaryBerne_PNAS16_SGOOP}.
While both SGOOP and VAC find a manifold of slow CVs by maximizing
the largest eigenvalues, SGOOP combines this principle with a maximum
Caliber based estimation of the transition or rate matrix, and is
thus applicable to enhanced-sampling simulations where dynamics are
not readily available. In the MMMK collection, Smith et al. develop
a multidimensional version of SGOOP by introducing conditional probability
factorization, and demonstrate its usefulness on the rare-event dissociation
pathway of benzene from Lysozyme \citep{SmithEtAl_JCPMSM19_SGOOP2}.

Paul et al. \citep{PaulEtAl_VAMP_dimred} generalize the idea of VAC
and TICA to nonequilibrium processes: They show that VAMP can be used
to variationally find slow CVs in systems that are driven by external
forces, such as an ion channel in an electrical field. Operationally,
the approach is as easy as TICA: time-correlation matrices between
features are computed and a singular value decomposition yields the
slow CVs. The MSMs estimated in this manifold reveal the circular
fluxes between long-lived states driven by the external potential
\citep{PaulEtAl_VAMP_dimred}. 

\paragraph{Estimating transition matrices and other quantities}

It is easy to show that the estimation of MSM transition matrices
by maximizing the Markov chain likelihood is statistically unbiased
if all simulations are in global equilibrium. However, MSMs are usually
estimated from short simulation trajectories that may be simulated
under equilibrium dynamics, but whose starting points do not start
from a global equilibrium distribution. Nüske et al. \citep{NueskeEtAl_JCP17_OOMMSM}
have derived the mathematical form of the MSM estimation error for
such data, and have proposed a reweighting method that allows the
user to estimate MSMs without this initial state bias. In the MMMK
collection, \citep{BacciEtAl_JCPMSM19_Initial} provide a new estimation
method for the same aim which is based on statistical resampling.
As always in machine learning, there is a trade-off between bias and
variance of estimators \citep{Vapnik_IEEE99_StatisticalLearningTheory}
that all these bias-reducing estimators must face. While most MSM
estimators have been developed with the aim of reducing the bias,
a systematic account for MSM estimators with an optimal bias-variance
trade-off is still an open issue for the future. 

As described above, slow CVs, transition rates and MSM transition
matrices can be viewed as the result of a variational optimization
process, e.g. using VAC, VAMP or SGOOP. From a mathematical point
of view, all methods which do this via some form of linear combination
of basis functions -- and this includes TICA and standard MSM transition
matrix estimators -- can also be described by the Galerkin approximation
framework \citep{SarichSchuette_MSMBook13}. The idea of the Galerkin
approach is as follows: we define basis functions -- the mean-free
feature functions in TICA or indicator functions denoting where Markov
states are in discrete MSMs -- and consider the projection of the
dynamics onto this basis set. The Galerkin approach then gives us
expression for dynamical quantities, such as the Markov propagator
eigenfunctions and its eigenvalues / relaxation rates, based on linear
combinations of these basis functions. While the Galerkin approach
is primarily a mathematical explanation of what happens in TICA or
MSM estimation algorithms, Thiede et al show in the MMMK collection
how it can be exploited and expanded to develop better MSM estimators,
and also obtain direct estimators for quantities that are usually
estimated via MSMs, such as committor functions \citep{ThiedeEtAl_JCPMSM19_Galerkin}.

As mentioned in the context of core-based MSMs, committors, mean first-passage
times (MFPTs), milestones and MSMs are deeply connected. In the MMMK
collection, Berezhkovskii and Szabo \citep{BerezhkovskiiSzabo_JCPMSM19_MSMcommitttor}
further our theoretical understanding of these relationships by showing
why exact MFPTs can be computed via a milestoning MSM and provide
a relationship between the equilibrium population of milestoning MSM
states and the committor functions.

Building upon previous work done on a variational framework for the
identification of Markovian transition states, \citep{MartiniEtAl_PRX17_VariationalTS},
Kells et al. \citep{KellsEtAl_JCPMSM19_MFPT} develop a variationally
optimal coarse-graining framework for MSM transition matrices that
has broad applicability and for time series analysis of large datasets
in general. They demonstrate that coarse-graining an MSM into two
or three states with this method has a simple physical interpretation
in terms of mean first passage times and fluxes between the coarse
grained states. Results are presented using both analytic test potentials
and MD simulations of pentalanine.

\paragraph{Markov model estimation with rare-events}

While MSMs effectively turn the problem of estimating molecular kinetics
and thermodynamics into an embarrassingly parallel process, estimating
a statistically precise or even connected MSM is still hampered by
sampling the rare transition events sufficiently often. A manifold
of MSM-based approaches have been proposed to address the sampling
problem, most prominently: 1) Adaptive sampling approaches, where
an MSM is used which starting points for new simulations are most
promising to discover new states or reduce statistical error \citep{Singhal_JCP07,HuangBowmanPande_PNAS09_AdaptiveSeeding,BowmanEnsignPande_JCTC2010_AdaptiveSampling,PretoClementi_PCCP14_AdaptiveSampling,DoerrDeFabritiis_JCTC14_OnTheFly,DoerrEtAl_JCTC16_HTMD,PlattnerEtAl_NatChem17_BarBar,HruskaEtAl_JCPMSM19_Adaptive},
and 2) multi-ensemble Markov modeling approaches, which estimate MSMs
with the aid of generalized ensemble simulations (multiples temperatures
or biases), in order to exploit expedited rare-event sampling \citep{WuMeyRostaNoe_JCP14_dTRAM,RostaHummer_DHAM,MeyWuNoe_xTRAM,WuEtAL_PNAS16_TRAM,StelzlHummer_JCTC17_KineticsREMD,PaulEtAl_PNAS17_Mdm2PMI}.
In the MMMK collection, several new methods are proposed to construct
MSMs without sampling the rare events by brute force.

Adaptive sampling is considered in Hruska et al. \citep{HruskaEtAl_JCPMSM19_Adaptive}.
While a variety of adaptive sampling methods have been developed before,
the authors conduct a systematic study of the effectiveness of different
adaptive sampling strategies on several fast folding proteins. \citep{HruskaEtAl_JCPMSM19_Adaptive}
provides theoretical limits for the adaptive sampling speed-up and
shows that different adaptive sampling strategies are optimal, depending
on sampling starts without prior knowledge of the metastable states,
or whether some states are already known and finding new ones is the
aim. 

By combining the maximum caliber approach \citep{Jaynes_ARPC80_MaxCal,PresseEtAl_RMP13_MaxCal}
with optimal transport theory, Dixit and Dill develop an approach
to approximate MSM rate matrices from short non-equilibrium simulations
\citep{DixitDill_JCPMSM19_MSMtransport}. Maximum caliber-based estimation
of MSMs is used in Meral et al. \citep{MeralEtAl_JCPMSM19_DNA} in
combination with enhanced sampling using well-tempered Metadynamics
\citep{LaioParrinello_PNAS99_12562,BarducciBussiParrinello_PRL08_WellTemperedMetadynamics}.
The authors apply their framework to the challenging problem of studying
the activation of a G Protein-Coupled Receptor (GPCR), here the $\mu$-opioid
receptor. They demonstrate that the caliber-derived transition rates
are in agreement with those obtained from adaptive sampling, suggesting
that the framework is of general usefulness. 

Another approach to avoid waiting for the rare events to happen is
to speed up sampling between known metastable states using transition
path methods, such as transition path sampling \citep{BolhuisChandlerDellagoGeissler_AnnuRevPhysChem02_TPS},
transition interface sampling (TIS) \citep{Du2011}, or Forward Flux
Sampling (FFS) \citep{AllenFrenkelWolde_JCP06_FFS}, and subsequently
constructing coarse-grained MSMs from the transition path statistics.
Recently developed software for transition path simulations facilitates
this task \citep{Swenson_JCTC19_OPS1}. In the MMMK collection, Qin
et al. \citep{QinEtAl_JCPMSM19_TIS} develop the reweighted partial
path (RPP) method approach which can efficiently reweight TIS or FFS
simulations in order to derive equilibrium distributions of states
or free energy profiles. 

Path-based sampling is also considered in Zhu et al. \citep{ZhuEtAl_JCPMSM19_TAPS}.
The authors develop a new path-searching method for connecting different
metastable states of biomolecules that employs ideas from the traveling-salesman
problem. Their TAPS algorithm outperforms the string method by 5 to
8 times for peptides in vacuum and solution, suggesting that it is
an efficient method to obtain initial pathways and intermediates that
facilitate the construction of MSMs and thereby full kinetics of complex
conformational changes. 

\paragraph{Clustering and coarse-graining}

A successful class of kinetic models are core-based MSMs, originally
proposed in \citep{BucheteHummer_JPCB08}. Core-based MSMs directly
go from a low-dimensional manifold of feature space to an MSM of few
metastable states, skipping over the traditional approach of clustering
that space into microstates and coarse-graining the microstate transition
matrix. The basic idea of core-based MSMs is to identify dynamical
cores -- the most densely populated regions of state space which
are parts of metastable states -- and estimate an MSM from the rare
transition paths between cores. Theoretically, core MSMs are closely
related to milestoning \citep{FaradjianElber_JCP04_Milestoning},
can be shown to approximate committor functions between metastable
states, which are in turn approximating the eigenfunctions of the
Markov propagator in metastable systems \citep{SchuetteEtAl_JCP11_Milestoning}.

A natural approach to identify cores are density-based clustering
algorithms \citep{LemkeKeller_JCP16_DensityBasedClustering,SittelStock_JCP16_PathBasedClustering}.
In the MMMK collection, Nagel et al. \citep{NagelEtAl_JCPMSM19_DynamicalCoring}
propose an extension of their previous density-based coring algorithm
\citep{SittelStock_JCP16_PathBasedClustering}, which avoids misclassification
of MD simulation frames to cores by requiring a minimum time spend
in a new core to qualify as a core transition. They demonstrate that
dynamical coring obtains better MSMs using alanine dipeptide dynamics
and Villin headpiece folding as examples. 

\paragraph{Nonequilibrium Markov models}

Deviations from equilibrium can come in different forms: An ion channel
in an electric field may be in steady-state, i.e. it has an unique
stationary distribution, but does not obey detailed balance. A spectroscopically
probed molecule may be subject to a period external force. When a
molecular system is expanded by pulling it with a nonequilibrium optical
tweezer experiment, even the dynamics themselves become time-dependent
and neither a stationary distribution exists nor detailed balance
is obeyed. These different degrees of nonequilibrium call for different
analysis methods that are only beginning to unfold now. VAMP-based
identification of the slow kinetics manifold for nonequilibrium has
been discussed above \citep{PaulEtAl_VAMP_dimred}.

In \citep{ReuterEtAl_JCPMSM19_Schur}, Reuter et al. generalize the
popular robust Perron Cluster Cluster Analysis (PCCA+) method for
coarse-graining transition matrices to obtain metastable states. Their
generalized method (G-PCCA) decomposes the MSM transition matrix with
a Schur decomposition instead of an eigenvalue decomposition, and
can obtain metastable states as well as slow cyclical processes from
transition matrices that do not obey detailed balance.

Knoch and Speck \citep{KnochSpeck_JCPMSM19_Noneq} develop a method
to construct MSMs for systems that are periodically driven, and illustrate
the method using a alanine dipeptide molecule that is exposed to a
periodic electric field.

\paragraph{Hidden Markov Models and experimental data}

Hidden Markov models (HMMs) are an alternative to MSMs and have been
used to obtain few-state kinetic models from MD data \citep{NoeEtAl_PMMHMM_JCP13,McGibbon_ICML14_HMM}.
They have also been used to extract kinetics directly from experimental
trajectories, such as single-molecule FRET or optical tweezer measurements,
which only track one or a few experimental observables over time instead
of the full configuration vector \citep{GebhardRief_PNAS10_AFMEnergyLandscapeProtein,PirchiHaran_NatureComms11_SingleMoleculeFRET,KellerNoe_JACS14_HMM-FRET}.
Similar methods have been used in order to analyze the kinetics from
short FRET trajectories of single molecules diffusing through a confocal
volume \citep{GopichSzabo_FRETCorrelation_JCP09,GopichSzabe_HMMFRET_JCP09}.
Much of MSM theory can be reused when dealing with HMMs, for example
in order to compute relaxation times and a hierarchical decomposition
of the system kinetics into metastable states with different lifetimes
\citep{KellerNoe_JACS14_HMM-FRET}.

In the MMMK collection, Jazani et al. develop a HMM which analyzes
fluorescence data from molecules diffusing through a single confocal
volume \citep{JazaniEtAl_JCPMSM19_Tracking}. Although the fluorescence
data stems from a single sensor -- in contrast to wide-field optical
microscopy -- \citep{JazaniEtAl_JCPMSM19_Tracking} shows that the
intensity fluctuations resulting from the fact that the non-homogeneity
of the confocal excitation volume bears information about the spatial
location of the molecule that can be exploited to reconstruct molecular
diffusion paths. 

\paragraph{Machine Learning}

Recently, the classical approach of constructing MSMs has been disrupted
by replacing the traditional estimation pipeline (see above) by VAMPnets,
where a deep neural network is trained with VAMP to map from high-dimensional
coordinate or feature space to a few-state MSM \citep{MardtEtAl_VAMPnets}.
Deep neural networks are now routinely used for several MSM-related
tasks, such as learning slow CVs or aiding rare event sampling \citep{RibeiroTiwary_JCP18_RAVE,ZhangYanNoe_ChemRxiv19_TALOS,Hernandez_PRE18_VTE,Jung2019,WangEtAl_NatComm19_PastFutureBottleneck}.

Another important machine-learning framework are kernel methods. Kernel
methods have been previously used for TICA \citep{HarmelingEtAl_NeurComput03_KernelTDSEP,SchwantesPande_JCTC15_kTICA},
and are also underlying the diffusion map approach that is a popular
MD analysis framework \citep{CoifmanLafon_PNAS05_DiffusionMaps,RohrdanzClementi_JCP134_DiffMaps}.
Following a similar approach as VAMPnets. Klus et al. develop a VAMP-optimal
kernel method \citep{KlusEtAl_JCPMSM19_Kernel} to estimate conformation
dynamics directly using a kernel function acting on molecular feature
space. They show that established linear models such as TICA and MSMs
are special cases of their kernel model and demonstrate the computation
of metastable states and kinetics for alanine dipeptide dynamics and
NTL9 protein folding. 

\paragraph{Software}

An important technology driving MSM method development, dissemination
and application is publicly available and software. Luckily, two large-scale
and widely used open-source packages, PyEMMA \citep{SchererEtAl_JCTC15_EMMA2}
and MSMbuilder \citep{HarriganEtAl_BJ17_MSMbuilder}, exist that implement
a wide range of MSM methods and welcome contributions from the community.

Recently, new software packages have added that publish additional
MSM methods and techniques, e.g. \citep{Swenson_JCTC19_OPS1,DeSancho_JCIM19_MasterMSM}.
In the MMMK collection, Porter et al. \citep{PorterEtAl_JCPMSM19_Enspara}
present the Enspara library which is geared towards scalability to
large data or models, i.e. MSMs with many states or from very large
datasets. Enspara includes parallelized implementations of computationally
intensive operations, and represents a flexible framework for MSM
construction and analysis.

\paragraph{Applications}

While MSMs and related techniques in the molecular sciences have been
primarily developed to study peptide and protein folding, they are
now used for a wide range of dynamical processes, including the study
of liquids, aggregation, and structural transitions in materials such
as alloys. The MMMK article collection is no exception and contains
MSM applications to various interesting molecular processes. In all
of these examples, the MSM framework reveals new physical or biological
insight by revealing structures and transition processes at an unprecedented
degree of detail.

Liquid water is a surprisingly rich and complex dynamical system.
Long-standing questions, for example, include which dynamical rearrangements
lead to the picosecond dynamics observed in spectroscopic data of
liquid water. The difficulty in answering such questions -- even
with accurate molecular models at hand -- lies in data analysis:
how can we define ``state space'' in a practical way and which molecular
features are suitable in a liquid of molecules that are diffusing
around, constantly switching between states that are identical up
to the exchange of labels. In the MMMK collection, Schulz et al. use
MSM methodology to pursue a detailed analysis of liquid water \citep{SchulzEtAl_JCPMSM19_DNA}.
They solve the permutation-invariance problem by considering each
water trimer as a subsystem in a 12-dimensional space defined by aligning
the coordinate system to one of the water molecules, and then perform
an MSM analysis using all water trimer trajectories of a solvent box
simulation in this space. The analysis suggests which exact transition
processes are observed by experiment and how elementary dynamical
processes, such as hydrogen-bond exchange in liquid water occur in
detail.

Gopich and Szabo \citep{GopichSzabo_JCPMSM19_TwoSite} work out a
detailed analysis of diffusion-limited kinetics of a ligand to a macromolecule
with two competing binding sites. Their results indicate that the
kinetics of such a system are surprisingly rich, the presence of the
second empty binding site can slow down binding to the first as a
result of competition, or it can speed up binding when populated as
direct transitions of ligands between the two binding sites are possible.

Shin and Kolomeisky employ MSM methods in order to model the kinetics
of a one-dimensional walker with conformational changes that affect
its transition probabilities \citep{ShinKolomeisky_JCPMSM19_Walker}.
Biological systems have many examples of such processes, such as the
dynamics of molecular motors along filaments, whose motion depends
on the current conformation of the motor protein. The authors derive
a phase diagram of such systems exhibiting several dynamical regimes
of the one-dimensional search process that are determined by the ratios
of the relevant length scales. 

Pinamonti et al. combine an advanced clustering technique with core-based
MSMs in order to analyze the process of RNA base fraying in detail
\citep{PinamontiEtAl_JCPMSM19_DNA}. The dynamics of four different
RNA duplexes are analyzed and an interesting interplay between the
equilibrium probability of intermediate states and the overall fraying
kinetics is described.

Chakraborty and Wales \citep{ChakrabortyWales_JCPMSM19_DPS} obtain
an MSM of the adenine-adenine RNA conformational switch using the
discrete path sampling technique (DPS) \citep{Wales_MolPhys100_3285}.
DPS allows the authors to probe very rare events, with interconversion
time scale here predicted to be in the range of minutes. Several competing
structures, separated by high barriers are found but the two main
energy funnels lead to the major and minor conformations known from
NMR experiments.

Similar issues with permutation invariances exist when studying aggregation
and self-assembly of many identical molecules. To this end, Sengupta
et al. \citep{SenguptaEtAl_JCPMSM19_MFPT} construct CVs from descriptors
that are invariant with respect to permutation of identical molecules.
Using these CVs, the authors construct MSMs to describe the aggregation
of a subsequence of Alzheimer\textquoteright s amyloid-$\beta$ peptide.
The results suggest that disordered and $\beta$-sheet oligomers do
not interconvert, and thus amyloid formation relies on having formed
ordered aggregates from the very beginning. 

\paragraph{Conclusion}

Markov modeling has come of age. In the molecular sciences, it has
grown from an activity practiced by a handful of groups to a technique
used by a large fraction -- if not the majority -- of MD simulation
groups. Markov modeling has also gone beyond molecular sciences and
found applications in other areas of dynamical systems. Recently it
has been found that key MSM techniques have evolved in parallel in
other fields under different names \citep{ZieheEtAl_JMLR04_FastTDSEP,SchmidSesterhenn_APS08_DMD,TuEtAl_JCD14_ExactDMD,WilliamsKevrekidisRowley_JNS15_EDMD},
and consolidating these efforts is fruitful for all these fields.

While basic aspects of Markov modeling, such as steps of the data
processing pipeline outlined in the beginning of this editorial are
now well established and largely under control, new research questions
have emerged, such as how to treat nonequilibrium processes, how to
deal with systems with permutation invariance such as liquid and membrane
systems, and how to exploit modern machine learning methods for molecular
thermodynamics and kinetics. The contributions of the MMMK collection
are a cross-section of this change.

An important driving force for the development of the field was, and
is, the availability of open-source well-maintained software. Currently
MSM softwares are primarily developed and maintained by individual
groups. We believe that a key to make this development sustainable
and maintainable -- and therefore to preserve the accumulated methodological
knowledge for the community -- is to move these softwares from single
groups to communities, or in other words dissociate them from individual
principle investigators, e.g. by merging packages from different groups.
The next years will be crucial in order to show whether this step
succeeds, and therefore whether MSM research can proceed at full steam.

\paragraph*{Acknowledgements}

We are grateful to the staff and editors of Journal of Chemical Physics
who proposed the MMMK collection and did the heavy lifting in collecting
and editing papers, especially Erinn Brigham, John Straub and Peter
Hamm. F.N. acknowledges funding from Deutsche Forschungsgemeinschaft
(CRC 1114, projects C03 and A04) and European Commission (ERC CoG
772230 ``ScaleCell''). E.R. acknowledges funding from EPSRC (EP/R013012/1,
EP/L027151/1, and EP/N020669/1) and European Commission (ERC StG 757850
``BioNet'').

\bibliographystyle{unsrtnat}
\bibliography{all,own,special_issue}

\begin{thebibliography}{126}
\providecommand{\natexlab}[1]{#1}
\providecommand{\url}[1]{\texttt{#1}}
\expandafter\ifx\csname urlstyle\endcsname\relax
  \providecommand{\doi}[1]{doi: #1}\else
  \providecommand{\doi}{doi: \begingroup \urlstyle{rm}\Url}\fi

\bibitem[Sch\"{u}tte et~al.(1999)Sch\"{u}tte, Fischer, Huisinga, and
  Deuflhard]{SchuetteFischerHuisingaDeuflhard_JCompPhys151_146}
C.~Sch\"{u}tte, A.~Fischer, W.~Huisinga, and P.~Deuflhard.
\newblock {A Direct Approach to Conformational Dynamics based on Hybrid Monte
  Carlo}.
\newblock \emph{J. Comput. Phys.}, 151:\penalty0 146--168, 1999.

\bibitem[Swope et~al.(2004)Swope, Pitera, and
  Suits]{SwopePiteraSuits_JPCB108_6571}
W.~C. Swope, J.~W. Pitera, and F.~Suits.
\newblock {Describing protein folding kinetics by molecular dynamics
  simulations: 1. Theory}.
\newblock \emph{J. Phys. Chem. B}, 108:\penalty0 6571--6581, 2004.

\bibitem[No{\'e} et~al.(2007)No{\'e}, Horenko, Sch{\"u}tte, and
  Smith]{NoeHorenkeSchutteSmith_JCP07_Metastability}
F.~No{\'e}, I.~Horenko, C.~Sch{\"u}tte, and J.~C. Smith.
\newblock {Hierarchical Analysis of Conformational Dynamics in Biomolecules:
  Transition Networks of Metastable States}.
\newblock \emph{J. Chem. Phys.}, 126:\penalty0 155102, 2007.

\bibitem[Chodera et~al.(2007)Chodera, Dill, Singhal, Pande, Swope, and
  Pitera]{ChoderaEtAl_JCP07}
J.~D. Chodera, K.~A. Dill, N.~Singhal, V.~S. Pande, W.~C. Swope, and J.~W.
  Pitera.
\newblock {Automatic discovery of metastable states for the construction of
  Markov models of macromolecular conformational dynamics}.
\newblock \emph{J. Chem. Phys.}, 126:\penalty0 155101, 2007.

\bibitem[Buchete and Hummer(2008)]{BucheteHummer_JPCB08}
N.~V. Buchete and G.~Hummer.
\newblock {Coarse Master Equations for Peptide Folding Dynamics}.
\newblock \emph{J. Phys. Chem. B}, 112:\penalty0 6057--6069, 2008.

\bibitem[Prinz et~al.(2011)Prinz, Wu, Sarich, Keller, Senne, Held, Chodera,
  Sch{\"u}tte, and No{\'e}]{PrinzEtAl_JCP10_MSM1}
J.-H. Prinz, H.~Wu, M.~Sarich, B.~G. Keller, M.~Senne, M.~Held, J.~D. Chodera,
  C.~Sch{\"u}tte, and F.~No{\'e}.
\newblock Markov models of molecular kinetics: Generation and validation.
\newblock \emph{J. Chem. Phys.}, 134:\penalty0 174105, 2011.

\bibitem[Bowman et~al.(2014)Bowman, Pande, and No{\'e}]{BowmanPandeNoe_MSMBook}
G.~R. Bowman, V.~S. Pande, and F.~No{\'e}, editors.
\newblock \emph{An Introduction to Markov State Models and Their Application to
  Long Timescale Molecular Simulation.}, volume 797 of \emph{Advances in
  Experimental Medicine and Biology}.
\newblock Springer Heidelberg, 2014.

\bibitem[Sch\"{u}tte and Sarich(2013)]{SarichSchuette_MSMBook13}
C.~Sch\"{u}tte and M.~Sarich.
\newblock \emph{Metastability and Markov State Models in Molecular Dynamics}.
\newblock Courant Lecture Notes. American Mathematical Society, 2013.

\bibitem[Chodera and No{\'e}(2014)]{ChoderaNoe_COSB14_MSMs}
J.~D. Chodera and F~No{\'e}.
\newblock Markov state models of biomolecular conformational dynamics.
\newblock \emph{Curr. Opin. Struc. Biol.}, 25:\penalty0 135--144, 2014.

\bibitem[Husic and Pande(2018)]{HusicPande_JACS18_MSMReview}
B.~E. Husic and V.~S. Pande.
\newblock Markov state models: From an art to a science.
\newblock \emph{J. Am. Chem. Soc.}, 140:\penalty0 2386--2396, 2018.

\bibitem[Scherer et~al.(2015)Scherer, Trendelkamp-Schroer, Paul,
  Perez-Hernandez, Hoffmann, Plattner, Wehmeyer, Prinz, and
  No{\'e}]{SchererEtAl_JCTC15_EMMA2}
M.~K. Scherer, B.~Trendelkamp-Schroer, F.~Paul, G.~Perez-Hernandez,
  M.~Hoffmann, N.~Plattner, C.~Wehmeyer, J.-H. Prinz, and F.~No{\'e}.
\newblock {PyEMMA 2: A software package for estimation, validation and analysis
  of Markov models}.
\newblock \emph{J. Chem. Theory Comput.}, 11:\penalty0 5525--5542, 2015.

\bibitem[Doerr et~al.(2016)Doerr, Harvey, No\'{e}, and
  Fabritiis]{DoerrEtAl_JCTC16_HTMD}
S.~Doerr, M.~J. Harvey, F.~No\'{e}, and G.~De Fabritiis.
\newblock {HTMD: High-Throughput Molecular Dynamics for Molecular Discovery}.
\newblock \emph{J. Chem. Theory Comput.}, 12:\penalty0 1845--1852, 2016.

\bibitem[Harrigan et~al.(2017)Harrigan, Sultan, Hern{\'a}ndez, Husic, Eastman,
  Schwantes, Beauchamp, McGibbon, and Pande]{HarriganEtAl_BJ17_MSMbuilder}
M.~P. Harrigan, M.~M. Sultan, C.~X. Hern{\'a}ndez, B.~E. Husic, P.~Eastman,
  C.~R. Schwantes, K.~A. Beauchamp, R.~T. McGibbon, and V.~S. Pande.
\newblock Msmbuilder: Statistical models for biomolecular dynamics.
\newblock \emph{Biophys J.}, 112:\penalty0 10--15, 2017.

\bibitem[Humphrey et~al.(1996)Humphrey, Dalke, and
  Schulten]{HumphreyDalkeSchulten_JMG96_VMD}
W.~Humphrey, A.~Dalke, and K.~Schulten.
\newblock Vmd - visual molecular dynamics.
\newblock \emph{J. Molec. Graphics}, 14:\penalty0 33--38, 1996.

\bibitem[McGibbon et~al.(2015)McGibbon, Beauchamp, Harrigan, Klein, Swails,
  Hern{\'a}ndez, Schwantes, Wang, Lane, and Pande]{McGibbon_BJ15_MDTraj}
R.~T. McGibbon, K.~A. Beauchamp, M.~P. Harrigan, C.~Klein, J.~M. Swails, C.~X.
  Hern{\'a}ndez, C.~R. Schwantes, L.~P. Wang, T.~J. Lane, and V.~S. Pande.
\newblock Mdtraj: A modern open library for the analysis of molecular dynamics
  trajectories.
\newblock \emph{Biophys J.}, 109:\penalty0 1528--1532, 2015.

\bibitem[No{\'e} and N{\"u}ske(2013)]{NoeNueske_MMS13_VariationalApproach}
F.~No{\'e} and F.~N{\"u}ske.
\newblock A variational approach to modeling slow processes in stochastic
  dynamical systems.
\newblock \emph{Multiscale Model. Simul.}, 11:\penalty0 635--655, 2013.

\bibitem[N{\"u}ske et~al.(2014)N{\"u}ske, Keller, P{\'e}rez-Hern{\'a}ndez, Mey,
  and No{\'e}]{NueskeEtAl_JCTC14_Variational}
F.~N{\"u}ske, B.~G. Keller, G.~P{\'e}rez-Hern{\'a}ndez, A.~S. J.~S. Mey, and
  F.~No{\'e}.
\newblock Variational approach to molecular kinetics.
\newblock \emph{J. Chem. Theory Comput.}, 10:\penalty0 1739--1752, 2014.

\bibitem[Perez-Hernandez et~al.(2013)Perez-Hernandez, Paul, Giorgino, {D
  Fabritiis}, and No{\'e}]{PerezEtAl_JCP13_TICA}
G.~Perez-Hernandez, F.~Paul, T.~Giorgino, G.~{D Fabritiis}, and Frank No{\'e}.
\newblock Identification of slow molecular order parameters for markov model
  construction.
\newblock \emph{J. Chem. Phys.}, 139:\penalty0 015102, 2013.

\bibitem[Schwantes and Pande(2013)]{SchwantesPande_JCTC13_TICA}
C.~R. Schwantes and V.~S. Pande.
\newblock Improvements in markov state model construction reveal many
  non-native interactions in the folding of ntl9.
\newblock \emph{J. Chem. Theory Comput.}, 9:\penalty0 2000--2009, 2013.

\bibitem[Molgedey and Schuster(1994)]{Molgedey_94}
L.~Molgedey and H.~G. Schuster.
\newblock Separation of a mixture of independent signals using time delayed
  correlations.
\newblock \emph{Phys. Rev. Lett.}, 72:\penalty0 3634--3637, 1994.
\newblock \doi{10.1103/PhysRevLett.72.3634}.

\bibitem[Ziehe and M\"uller(1998)]{ZieheMueller_ICANN98_TDSEP}
A.~Ziehe and K.-R. M\"uller.
\newblock {TDSEP} - an efficient algorithm for blind separation using time
  structure.
\newblock In \emph{{ICANN} 98}, pages 675--680. Springer Science and Business
  Media, 1998.

\bibitem[Mezi\'{c}(2005)]{Mezic_NonlinDyn05_Koopman}
I.~Mezi\'{c}.
\newblock Spectral properties of dynamical systems, model reduction and
  decompositions.
\newblock \emph{Nonlinear Dynam.}, 41:\penalty0 309--325, 2005.

\bibitem[Schmid and Sesterhenn(2008)]{SchmidSesterhenn_APS08_DMD}
P.~J. Schmid and J.~Sesterhenn.
\newblock Dynamic mode decomposition of numerical and experimental data.
\newblock In \emph{61st Annual Meeting of the APS Division of Fluid Dynamics.
  American Physical Society}, 2008.

\bibitem[Williams et~al.(2015)Williams, Kevrekidis, and
  Rowley]{WilliamsKevrekidisRowley_JNS15_EDMD}
M.~O. Williams, I.~G. Kevrekidis, and C.~W. Rowley.
\newblock A data-driven approximation of the koopman operator: Extending
  dynamic mode decomposition.
\newblock \emph{J. Nonlinear Sci.}, 25:\penalty0 1307--1346, 2015.

\bibitem[Tu et~al.(2014)Tu, Rowley, Luchtenburg, Brunton, and
  Kutz]{TuEtAl_JCD14_ExactDMD}
J.~H. Tu, C.~W. Rowley, D.~M. Luchtenburg, S.~L. Brunton, and J.~N. Kutz.
\newblock On dynamic mode decomposition: Theory and applications.
\newblock \emph{J. Comput. Dyn.}, 1\penalty0 (2):\penalty0 391--421, dec 2014.
\newblock \doi{10.3934/jcd.2014.1.391}.

\bibitem[No{\'e} and Clementi(2017)]{NoeClementi_COSB17_SlowCVs}
F.~No{\'e} and C.~Clementi.
\newblock Collective variables for the study of long-time kinetics from
  molecular trajectories: theory and methods.
\newblock \emph{Curr. Opin. Struc. Biol.}, 43:\penalty0 141--147, 2017.

\bibitem[No\'{e} and Clementi(2015)]{NoeClementi_JCTC15_KineticMap}
F.~No\'{e} and C.~Clementi.
\newblock Kinetic distance and kinetic maps from molecular dynamics simulation.
\newblock \emph{J. Chem. Theory Comput.}, 11:\penalty0 5002--5011, 2015.

\bibitem[No\'{e} et~al.(2016)No\'{e}, Banisch, and
  Clementi]{NoeClementi_JCTC16_KineticMap2}
F.~No\'{e}, R.~Banisch, and C.~Clementi.
\newblock Commute maps: separating slowly-mixing molecular configurations for
  kinetic modeling.
\newblock \emph{J. Chem. Theory Comput.}, 12:\penalty0 5620--5630, 2016.

\bibitem[Husic and Pande(2017)]{HusicPande_JCTC17_Ward}
B.~E. Husic and V.~S. Pande.
\newblock Ward clustering improves cross-validated markov state models of
  protein folding.
\newblock \emph{J. Chem. Theo. Comp.}, 13:\penalty0 963--967, 2017.

\bibitem[Sheong et~al.(2015)Sheong, Silva, Meng, Zhao, and
  Huang]{SheongEtAl_JCTC15_APM}
F.~K. Sheong, D.-A. Silva, L.~Meng, Y.~Zhao, and X.~Huang.
\newblock {Automatic State Partitioning for Multibody Systems (APM): An
  Efficient Algorithm for Constructing Markov State Models To Elucidate
  Conformational Dynamics of Multibody Systems}.
\newblock \emph{J. Chem. Theory Comput.}, 11:\penalty0 17--27, 2015.

\bibitem[Wu and No\'{e}(2015)]{WuNoe_JCP15_GMTM}
H.~Wu and F.~No\'{e}.
\newblock Gaussian markov transition models of molecular kinetics.
\newblock \emph{J. Chem. Phys.}, 142:\penalty0 084104, 2015.

\bibitem[Harrigan and Pande(2017)]{HarriganPande_bioRxiv17_LandmarkTICA}
M.~P. Harrigan and V.~S. Pande.
\newblock Landmark kernel tica for conformational dynamics.
\newblock \emph{bioRxiv, https://doi.org/10.1101/123752}, 2017.

\bibitem[Weber et~al.(2017)Weber, Fackeldey, and
  Sch{\"u}tte]{WeberFackeldeySchuette_JCP17_SetfreeMSM}
M.~Weber, K.~Fackeldey, and C.~Sch{\"u}tte.
\newblock Set-free markov state model building.
\newblock \emph{J. Chem. Phys.}, 146:\penalty0 124133,, 2017.

\bibitem[Bowman et~al.(2009)Bowman, Beauchamp, Boxer, and
  Pande]{Bowman_JCP09_Villin}
G.~R. Bowman, K.~A. Beauchamp, G.~Boxer, and V.~S. Pande.
\newblock {Progress and challenges in the automated construction of Markov
  state models for full protein systems.}
\newblock \emph{J. Chem. Phys.}, 131:\penalty0 124101, 2009.

\bibitem[Trendelkamp-Schroer et~al.(2015)Trendelkamp-Schroer, Wu, Paul, and
  No\'{e}]{TrendelkampSchroerEtAl_InPrep_revMSM}
B.~Trendelkamp-Schroer, H.~Wu, F.~Paul, and F.~No\'{e}.
\newblock Estimation and uncertainty of reversible markov models.
\newblock \emph{J. Chem. Phys.}, 143:\penalty0 174101, 2015.

\bibitem[Deuflhard and Weber(2005)]{DeuflhardWeber_LAA05_PCCA+}
P.~Deuflhard and M.~Weber.
\newblock Robust perron cluster analysis in conformation dynamics.
\newblock In M.~Dellnitz, S.~Kirkland, M.~Neumann, and C.~Sch{\"u}tte, editors,
  \emph{Linear Algebra Appl.}, volume 398C, pages 161--184. Elsevier, New York,
  2005.

\bibitem[Kube and Weber(2007)]{KubeWeber_JCP07_CoarseGraining}
S.~Kube and M.~Weber.
\newblock {A coarse graining method for the identification of transition rates
  between molecular conformations}.
\newblock \emph{J. Chem. Phys.}, 126:\penalty0 024103, 2007.

\bibitem[Yao et~al.(2013)Yao, Cui, Bowman, Silva, Sun, and
  Huang]{YaoHuang_JCP13_Nystrom}
Y.~Yao, R.~Z. Cui, G.~R. Bowman, D.-A. Silva, J.~Sun, and X.~Huang.
\newblock Hierarchical nystr{\"o}m methods for constructing markov state models
  for conformational dynamics.
\newblock \emph{J. Chem. Phys.}, 138:\penalty0 174106, 2013.

\bibitem[Fackeldey and Weber(2017)]{FackeldeyWeber_WIAS17_GenPCCA}
K.~Fackeldey and M.~Weber.
\newblock Genpcca -- markov state models for non-equilibrium steady states.
\newblock \emph{WIAS Report}, 29:\penalty0 70--80, 2017.

\bibitem[Gerber and Horenko(2017)]{GerberHorenko_PNAS17_Categorial}
S.~Gerber and I.~Horenko.
\newblock Toward a direct and scalable identification of reduced models for
  categorical processes.
\newblock \emph{Proc. Natl. Acad. Sci. USA}, 114:\penalty0 4863--4868, 2017.

\bibitem[Hummer and Szabo(2015)]{HummerSzabo_JPCB15_CoarseGraining}
G.~Hummer and A.~Szabo.
\newblock Optimal dimensionality reduction of multistate kinetic and
  markov-state models.
\newblock \emph{J. Phys. Chem. B}, 119:\penalty0 9029--9037, 2015.

\bibitem[Orioli and Faccioli(2016)]{OrioliFaccioli_JCP16_CoarseMSM}
S.~Orioli and P.~Faccioli.
\newblock Dimensional reduction of markov state models from renormalization
  group theory.
\newblock \emph{J. Chem. Phys.}, 145:\penalty0 124120, 2016.

\bibitem[No{\'e} et~al.(2013)No{\'e}, Wu, Prinz, and
  Plattner]{NoeEtAl_PMMHMM_JCP13}
F.~No{\'e}, H.~Wu, J.-H. Prinz, and N.~Plattner.
\newblock Projected and hidden markov models for calculating kinetics and
  metastable states of complex molecules.
\newblock \emph{J. Chem. Phys.}, 139:\penalty0 184114, 2013.

\bibitem[Sch{\"{u}}tte and Huisinga(2003)]{Schuette_HandbookNumAnal_2003}
C.~Sch{\"{u}}tte and W.~Huisinga.
\newblock {Biomolecular conformations can be identified as metastable sets of
  molecular dynamics}.
\newblock In P.~G. Ciaret and J.~L. Lions, editors, \emph{Handbook of Numerical
  Analysis}, volume X: Computational Chemistry, pages 699--744. North-Holland,
  2003.

\bibitem[Bittracher et~al.(2018)Bittracher, Koltai, Klus, Banisch, Dellnitz,
  and Sch{\"u}tte]{BittracherEtAl_arXiv2017_TransitionManifols}
A.~Bittracher, P.~Koltai, S.~Klus, R.~Banisch, M.~Dellnitz, and Ch.
  Sch{\"u}tte.
\newblock Transition manifolds of complex metastable systems: Theory and
  data-driven computation of effective dynamics.
\newblock \emph{J. Nonlinear Sci.}, 28:\penalty0 471--512, 2018.

\bibitem[McGibbon and Pande(2015)]{McGibbonPande_JCP15_CrossValidation}
R.~T. McGibbon and V.~S. Pande.
\newblock Variational cross-validation of slow dynamical modes in molecular
  kinetics.
\newblock \emph{J. Chem. Phys.}, 142:\penalty0 124105, 2015.

\bibitem[Koltai et~al.(2018)Koltai, Wu, No{\'e}, and
  Sch{\"u}tte]{KoltaiEtAl_Computation18_NonrevMSM}
P.~Koltai, H.~Wu, F.~No{\'e}, and C.~Sch{\"u}tte.
\newblock Optimal data-driven estimation of generalized markov state models for
  non-equilibrium dynamics.
\newblock \emph{Computation}, 6:\penalty0 22, 2018.

\bibitem[Reuter et~al.(2019)Reuter, Fackeldey, and
  Weber]{ReuterEtAl_JCPMSM19_Schur}
B.~Reuter, K.~Fackeldey, and M.~Weber.
\newblock Generalized markov modeling of nonreversible molecular kinetics.
\newblock \emph{J. Chem. Phys.}, 150:\penalty0 174103, 2019.

\bibitem[Knoch and Speck(2019)]{KnochSpeck_JCPMSM19_Noneq}
F.~Knoch and T.~Speck.
\newblock Non-equilibrium markov state modeling of periodically driven
  biomolecules.
\newblock \emph{J. Chem. Phys.}, 150:\penalty0 054103, 2019.

\bibitem[Knoch and Speck(2015)]{KnochSpeck_NJP15_NoneqMSM}
F.~Knoch and T.~Speck.
\newblock Cycle representatives for the coarse-graining of systems driven into
  a non-equilibrium steady state.
\newblock \emph{New J. Phys.}, 17:\penalty0 115004, 2015.

\bibitem[Wu and No\'{e}(2017)]{WuNoe_VAMP}
H.~Wu and F.~No\'{e}.
\newblock Variational approach for learning markov processes from time series
  data.
\newblock \emph{arXiv:1707.04659}, 2017.

\bibitem[Mardt et~al.(2018)Mardt, Pasquali, Wu, and
  No\'{e}]{MardtEtAl_VAMPnets}
A.~Mardt, L.~Pasquali, H.~Wu, and F.~No\'{e}.
\newblock Vampnets: Deep learning of molecular kinetics.
\newblock \emph{Nat. Commun.}, 9:\penalty0 5, 2018.

\bibitem[Jarzynski(1997)]{Jarzynski_PRL97}
C.~Jarzynski.
\newblock Nonequilibrium equality for free energy differences.
\newblock \emph{Phys. Rev. Lett.}, 78:\penalty0 2690, 1997.

\bibitem[Vitalini et~al.(2015)Vitalini, No\'{e}, and
  Keller]{VitaliniNoeKeller_JCTC15_BasisSet}
F.~Vitalini, F.~No\'{e}, and B.~G. Keller.
\newblock A basis set for peptides for the variational approach to
  conformational kinetics.
\newblock \emph{J. Chem. Theory Comput.}, 11:\penalty0 3992--4004, 2015.

\bibitem[Scherer et~al.(2019)Scherer, Husic, Hoffmann, F.~Paul, and
  No{\'e}]{SchererEtAl_JCPMSM19_VAMPFeatures}
M.~K. Scherer, B.~E. Husic, M.~Hoffmann, H.~Wu F.~Paul, and F.~No{\'e}.
\newblock Variational selection of features for molecular kinetics.
\newblock \emph{J. Chem. Phys.}, 150:\penalty0 194108, 2019.

\bibitem[Lindorff-Larsen et~al.(2011)Lindorff-Larsen, Piana, Dror, and
  Shaw]{LindorffLarsenEtAl_Science11_AntonFolding}
K.~Lindorff-Larsen, S.~Piana, R.~O. Dror, and D.~E. Shaw.
\newblock How fast-folding proteins fold.
\newblock \emph{Science}, 334:\penalty0 517--520, 2011.

\bibitem[Naritomi and Fuchigami(2011)]{NaritomiFuchigami_JCP11_TICA}
Y.~Naritomi and S.~Fuchigami.
\newblock Slow dynamics in protein fluctuations revealed by time-structure
  based independent component analysis: The case of domain motions.
\newblock \emph{J. Chem. Phys.}, 134\penalty0 (6):\penalty0 065101, 2011.

\bibitem[Karasawa et~al.(2019)Karasawa, Mitsutake, and
  Takano]{KarasawaEtAl_JCPMSM19_RMA}
N.~Karasawa, A.~Mitsutake, and H.~Takano.
\newblock Identification of slow relaxation modes in a protein trimer via
  positive definite relaxation mode analysis.
\newblock \emph{J. Chem. Phys.}, 150:\penalty0 084113, 2019.

\bibitem[Takano and
  Miyashita(1995)]{TakanoMiyashita_JPSJ95_RelaxationModeAnalysis}
H.~Takano and S.~Miyashita.
\newblock Relaxation modes in random spin systems.
\newblock \emph{J. Phys. Soc. Jpn.}, 64:\penalty0 3688--3698, 1995.

\bibitem[Tiwary and Berne(2016)]{TiwaryBerne_PNAS16_SGOOP}
P.~Tiwary and B.~J. Berne.
\newblock Spectral gap optimization of order parameters for sampling complex
  molecular systems.
\newblock \emph{Proc. Natl. Acad. Sci. USA}, 113:\penalty0 2839--2844, 2016.

\bibitem[Smith et~al.(2019)Smith, Pramanik, Tsai, and
  Tiwary]{SmithEtAl_JCPMSM19_SGOOP2}
Z.~Smith, D.~Pramanik, S.-T. Tsai, and P.~Tiwary.
\newblock Multi-dimensional spectral gap optimization of order parameters
  (sgoop) through conditional probability factorization.
\newblock \emph{J. Chem. Phys.}, 150:\penalty0 234105, 2019.

\bibitem[Paul et~al.(2018)Paul, Wu, Vossel, {de Groot}, and
  No{\'e}]{PaulEtAl_VAMP_dimred}
F.~Paul, H.~Wu, M.~Vossel, B.~L. {de Groot}, and F.~No{\'e}.
\newblock Identification of kinetic order parameters for non-equilibrium
  dynamics.
\newblock \emph{J. Chem. Phys.}, MMMK:\penalty0 164120, 2018.

\bibitem[N{\"u}ske et~al.(2017)N{\"u}ske, Wu, Wehmeyer, Clementi, and
  No{\'e}]{NueskeEtAl_JCP17_OOMMSM}
F.~N{\"u}ske, H.~Wu, C.~Wehmeyer, C.~Clementi, and F.~No{\'e}.
\newblock Markov state models from short non-equilibrium simulations - analysis
  and correction of estimation bias.
\newblock \emph{J. Chem. Phys.}, 146:\penalty0 094104, 2017.

\bibitem[Bacci et~al.(2019)Bacci, Caflisch, and
  Vitalis]{BacciEtAl_JCPMSM19_Initial}
M.~Bacci, A.~Caflisch, and A.~Vitalis.
\newblock On the removal of initial state bias from simulation data.
\newblock \emph{J. Chem. Phys.}, 150:\penalty0 104105, 2019.

\bibitem[Vapnik(1999)]{Vapnik_IEEE99_StatisticalLearningTheory}
V.~N. Vapnik.
\newblock An overview of statistical learning theory.
\newblock \emph{IEEE Trans. Neur. Net.}, 10, 1999.

\bibitem[Thiede et~al.(2019)Thiede, Giannakis, Dinner, and
  Weare]{ThiedeEtAl_JCPMSM19_Galerkin}
E.~H. Thiede, D.~Giannakis, A.~R. Dinner, and J.~Weare.
\newblock Galerkin approximation of dynamical quantities using trajectory data.
\newblock \emph{J. Chem. Phys.}, 150:\penalty0 244111, 2019.

\bibitem[Berezhkovskii and
  Szabo(2019)]{BerezhkovskiiSzabo_JCPMSM19_MSMcommitttor}
A.~M. Berezhkovskii and A.~Szabo.
\newblock Committors, first-passage times, fluxes, markov states, milestones,
  and all that.
\newblock \emph{J. Chem. Phys.}, 150:\penalty0 054106, 2019.

\bibitem[Martini et~al.(2017)Martini, Kells, Covino, Hummer, Buchete, and
  Rosta]{MartiniEtAl_PRX17_VariationalTS}
L.~Martini, A.~Kells, R.~Covino, G.~Hummer, N.-V. Buchete, and E.~Rosta.
\newblock Variational identification of markovian transition states.
\newblock \emph{Phys. Rev. X}, 7:\penalty0 031060, 2017.

\bibitem[Kells et~al.(2019)Kells, Mih{\'a}lka, Annibale, and
  Rosta]{KellsEtAl_JCPMSM19_MFPT}
A.~Kells, Z.~{\'E}. Mih{\'a}lka, A.~Annibale, and E.~Rosta.
\newblock Mean first passage times in variational coarse graining using markov
  state models.
\newblock \emph{J. Chem. Phys.}, 150:\penalty0 134107, 2019.

\bibitem[Hinrichs and Pande(2007)]{Singhal_JCP07}
N.~S. Hinrichs and V.~S. Pande.
\newblock {Calculation of the distribution of eigenvalues and eigenvectors in
  Markovian state models for molecular dynamics}.
\newblock \emph{J. Chem. Phys.}, 126:\penalty0 244101, 2007.

\bibitem[Huang et~al.(2009)Huang, Bowman, Bacallado, and
  Pande]{HuangBowmanPande_PNAS09_AdaptiveSeeding}
X.~Huang, G.~R. Bowman, S.~Bacallado, and V.~S. Pande.
\newblock {Rapid equilibrium sampling initiated from nonequilibrium data}.
\newblock \emph{Proc. Natl. Acad. Sci. USA}, 106\penalty0 (47):\penalty0
  19765--19769, 2009.

\bibitem[Bowman et~al.(2010)Bowman, Ensign, and
  Pande]{BowmanEnsignPande_JCTC2010_AdaptiveSampling}
G.~R. Bowman, D.~L. Ensign, and V.~S. Pande.
\newblock {Enhanced Modeling via Network Theory: Adaptive Sampling of Markov
  State Models}.
\newblock \emph{J. Chem. Theory Comput.}, 6\penalty0 (3):\penalty0 787--794,
  2010.
\newblock \doi{10.1021/ct900620b}.

\bibitem[Preto and Clementi(2014)]{PretoClementi_PCCP14_AdaptiveSampling}
J.~Preto and C.~Clementi.
\newblock Fast recovery of free energy landscapes via diffusion-map-directed
  molecular dynamics.
\newblock \emph{Phys. Chem. Chem. Phys.}, 16:\penalty0 19181--19191, 2014.

\bibitem[Doerr and Fabritiis(2014)]{DoerrDeFabritiis_JCTC14_OnTheFly}
S.~Doerr and G.~De Fabritiis.
\newblock On-the-fly learning and sampling of ligand binding by high-throughput
  molecular simulations.
\newblock \emph{J. Chem. Theory Comput.}, 10:\penalty0 2064--2069, 2014.

\bibitem[Plattner et~al.(2017)Plattner, Doerr, Fabritiis, and
  No{\'e}]{PlattnerEtAl_NatChem17_BarBar}
N.~Plattner, S.~Doerr, G.~De Fabritiis, and F.~No{\'e}.
\newblock Protein-protein association and binding mechanism resolved in atomic
  detail.
\newblock \emph{Nat. Chem.}, 9:\penalty0 1005--1011, 2017.

\bibitem[Hruska et~al.(2019)Hruska, Abella, N{\"u}ske, Kavraki, and
  Clementi]{HruskaEtAl_JCPMSM19_Adaptive}
E.~Hruska, J.~R. Abella, F.~N{\"u}ske, L.~E. Kavraki, and C.~Clementi.
\newblock Quantitative comparison of adaptive sampling methods for protein
  dynamics.
\newblock \emph{J. Chem. Phys.}, 150:\penalty0 244119, 2019.

\bibitem[Wu et~al.(2014)Wu, Mey, Rosta, and No\'{e}]{WuMeyRostaNoe_JCP14_dTRAM}
H.~Wu, A.~S. J.~S. Mey, E.~Rosta, and F.~No\'{e}.
\newblock Statistically optimal analysis of state-discretized trajectory data
  from multiple thermodynamic states.
\newblock \emph{J. Chem. Phys.}, 141:\penalty0 214106, 2014.

\bibitem[Rosta and Hummer(2015)]{RostaHummer_DHAM}
E.~Rosta and G.~Hummer.
\newblock Free energies from dynamic weighted histogram analysis using unbiased
  markov state model.
\newblock \emph{J. Chem. Theory Comput.}, 11:\penalty0 276--285, 2015.

\bibitem[Mey et~al.(2014)Mey, Wu, and No\'{e}]{MeyWuNoe_xTRAM}
A.~S. J.~S. Mey, H.~Wu, and F.~No\'{e}.
\newblock {xTRAM: Estimating equilibrium expectations from time-correlated
  simulation data at multiple thermodynamic states}.
\newblock \emph{Phys. Rev. X}, 4:\penalty0 041018, 2014.

\bibitem[Wu et~al.(2016)Wu, Paul, Wehmeyer, and No{\'e}]{WuEtAL_PNAS16_TRAM}
H.~Wu, F.~Paul, C.~Wehmeyer, and F.~No{\'e}.
\newblock Multiensemble markov models of molecular thermodynamics and kinetics.
\newblock \emph{Proc. Natl. Acad. Sci. USA}, 113:\penalty0 E3221--E3230, 2016.

\bibitem[Stelzl and Hummer(DOI:
  10.1021/acs.jctc.7b00372)]{StelzlHummer_JCTC17_KineticsREMD}
L.~S. Stelzl and G.~Hummer.
\newblock Kinetics from replica exchange molecular dynamics simulations.
\newblock \emph{J. Chem. Theory Comput.}, DOI: 10.1021/acs.jctc.7b00372.

\bibitem[Paul et~al.(2017)Paul, Wehmeyer, Abualrous, Wu, Crabtree,
  Sch{\"o}neberg, Clarke, Freund, Weikl, and No{\'e}]{PaulEtAl_PNAS17_Mdm2PMI}
F.~Paul, C.~Wehmeyer, E.~T. Abualrous, H.~Wu, M.~D. Crabtree,
  J.~Sch{\"o}neberg, J.~Clarke, C.~Freund, T.~R. Weikl, and F.~No{\'e}.
\newblock Protein-ligand kinetics on the seconds timescale from atomistic
  simulations.
\newblock \emph{Nat. Commun.}, 8:\penalty0 1095, 2017.

\bibitem[Jaynes(1980)]{Jaynes_ARPC80_MaxCal}
E.~T. Jaynes.
\newblock The minimum entropy production principle.
\newblock \emph{Ann. Rev. Phys. Chem.}, 31:\penalty0 579, 1980.

\bibitem[Press{\'e} et~al.(2013)Press{\'e}, Ghosh, Lee, and
  Dill]{PresseEtAl_RMP13_MaxCal}
S.~Press{\'e}, K.~Ghosh, J.~Lee, and K.~A. Dill.
\newblock Principles of maximum entropy and maximum caliber in statistical
  physics.
\newblock \emph{Rev. Mod. Phys.}, 85:\penalty0 1115, 2013.

\bibitem[Dixit and Dill(2019)]{DixitDill_JCPMSM19_MSMtransport}
P.~D. Dixit and K.~A. Dill.
\newblock Building markov state models using optimal transport theory.
\newblock \emph{J. Chem. Phys.}, 150:\penalty0 054105, 2019.

\bibitem[Meral et~al.(2019)Meral, Provasi, and
  Filizola]{MeralEtAl_JCPMSM19_DNA}
D.~Meral, D.~Provasi, and M.~Filizola.
\newblock An efficient strategy to estimate thermodynamics and kinetics of g
  protein-coupled receptor activation using metadynamics and maximum caliber.
\newblock \emph{J. Chem. Phys.}, 150:\penalty0 224101, 2019.

\bibitem[Laio and Parrinello(2002)]{LaioParrinello_PNAS99_12562}
A.~Laio and M.~Parrinello.
\newblock {Escaping free energy minima}.
\newblock \emph{Proc. Natl. Acad. Sci. USA}, 99:\penalty0 12562--12566, 2002.

\bibitem[Barducci et~al.(2008)Barducci, Bussi, and
  Parrinello]{BarducciBussiParrinello_PRL08_WellTemperedMetadynamics}
A.~Barducci, G.~Bussi, and M.~Parrinello.
\newblock Well-tempered metadynamics: A smoothly converging and tunable
  free-energy method.
\newblock \emph{Phys. Rev. Lett.}, 100:\penalty0 020603, 2008.

\bibitem[Bolhuis et~al.(2002)Bolhuis, Chandler, Dellago, and
  Geissler]{BolhuisChandlerDellagoGeissler_AnnuRevPhysChem02_TPS}
P.~G. Bolhuis, D.~Chandler, C.~Dellago, and P.~L. Geissler.
\newblock {Transition path sampling: throwing ropes over rough mountain passes,
  in the dark.}
\newblock \emph{Annu. Rev. Phys. Chem.}, 53\penalty0 (1):\penalty0 291--318,
  2002.

\bibitem[Du et~al.(2011)Du, Marino, and Bolhuis]{Du2011}
W.-N. Du, K.~A. Marino, and P.~G. Bolhuis.
\newblock Multiple state transition interface sampling of alanine dipeptide in
  explicit solvent.
\newblock \emph{J. Chem. Phys.}, 135\penalty0 (14):\penalty0 145102, 2011.

\bibitem[Allen et~al.(2006)Allen, Frenkel, and ten
  Wolde]{AllenFrenkelWolde_JCP06_FFS}
R.~J. Allen, D.~Frenkel, and P.~R. ten Wolde.
\newblock Simulating rare events in equilibrium or nonequilibrium stochastic
  systems.
\newblock \emph{J. Chem. Phys.}, 124:\penalty0 024102, 2006.

\bibitem[Swenson et~al.(2019)Swenson, Prinz, No{\'e}, Chodera, and
  Bolhuis]{Swenson_JCTC19_OPS1}
D.~W.~H. Swenson, J.-H. Prinz, F.~No{\'e}, J.~D. Chodera, and P.~G. Bolhuis.
\newblock Openpathsampling: A flexible, open framework for path sampling
  simulations. 1. basics.
\newblock \emph{J. Chem. Theory Comput.}, 15:\penalty0 813, 2019.

\bibitem[Qin et~al.(2019)Qin, Dellago, and Kozeschnik]{QinEtAl_JCPMSM19_TIS}
L.~Qin, C.~Dellago, and E.~Kozeschnik.
\newblock An efficient method to reconstruct free energy profiles for diffusive
  processes in transition interface sampling and forward flux sampling
  simulations.
\newblock \emph{J. Chem. Phys.}, 150:\penalty0 094114, 2019.

\bibitem[Zhu et~al.(2019)Zhu, Sheong, Cao, Liu, Unarta, and
  Huang]{ZhuEtAl_JCPMSM19_TAPS}
L.~Zhu, F.~K. Sheong, S.~Cao, S.~Liu, I.~C. Unarta, and X.~Huang.
\newblock Taps: A traveling-salesman based automated path searching method for
  functional conformational changes of biological macromolecules.
\newblock \emph{J. Chem. Phys.}, 150:\penalty0 124105, 2019.

\bibitem[Faradjian and Elber(2004)]{FaradjianElber_JCP04_Milestoning}
A.~K. Faradjian and R.~Elber.
\newblock Computing time scales from reaction coordinates by milestoning.
\newblock \emph{J. Chem. Phys.}, 120:\penalty0 10880, 2004.

\bibitem[Sch{\"u}tte et~al.(2011)Sch{\"u}tte, No{\'e}, Lu, Sarich, and
  Vanden-Eijnden]{SchuetteEtAl_JCP11_Milestoning}
C.~Sch{\"u}tte, F.~No{\'e}, J.~Lu, M.~Sarich, and E.~Vanden-Eijnden.
\newblock Markov state models based on milestoning.
\newblock \emph{J. Chem. Phys.}, 134:\penalty0 204105, 2011.

\bibitem[Lemke and Keller(2016)]{LemkeKeller_JCP16_DensityBasedClustering}
O.~Lemke and B.~G. Keller.
\newblock Density-based cluster algorithms for the identification of core sets.
\newblock \emph{J Chem Phys.}, 145:\penalty0 164104, 2016.

\bibitem[Sittel and Stock(2016)]{SittelStock_JCP16_PathBasedClustering}
F.~Sittel and G.~Stock.
\newblock Robust density-based clustering to identify metastable conformational
  states of proteins.
\newblock \emph{J. Chem. Theory Comput.}, 12:\penalty0 2426, 2016.

\bibitem[Nagel et~al.(2019)Nagel, Weber, Lickert, and
  Stock]{NagelEtAl_JCPMSM19_DynamicalCoring}
D.~Nagel, A.~Weber, B.~Lickert, and G.~Stock.
\newblock Dynamical coring of markov state models.
\newblock \emph{J. Chem. Phys.}, 150:\penalty0 094111, 2019.

\bibitem[McGibbon et~al.(2014)McGibbon, Ramsundar, Sultan, Kiss, and
  Pande]{McGibbon_ICML14_HMM}
R.~T. McGibbon, B.~Ramsundar, M.~M. Sultan, G.~Kiss, and V.~S. Pande.
\newblock Understanding protein dynamics with l1-regularized reversible hidden
  markov models.
\newblock In \emph{Proc. Int. Conf. Mach. Learn.}, volume~31, pages 1197--1205,
  2014.

\bibitem[Gebhardt et~al.(2010)Gebhardt, Bornschl\"{o}gl, and
  Rief]{GebhardRief_PNAS10_AFMEnergyLandscapeProtein}
J.~C. Gebhardt, T.~Bornschl\"{o}gl, and M.~Rief.
\newblock {Full distance-resolved folding energy landscape of one single
  protein molecule}.
\newblock \emph{Proc. Natl. Acad. Sci. USA}, 107\penalty0 (5):\penalty0
  2013--2018, 2010.

\bibitem[Pirchi et~al.(2011)Pirchi, Ziv, Riven, Cohen, Zohar, Barak, and
  Haran]{PirchiHaran_NatureComms11_SingleMoleculeFRET}
M.~Pirchi, G.~Ziv, I.~Riven, S.~S. Cohen, N.~Zohar, Y.~Barak, and G.~Haran.
\newblock Single-molecule fluorescence spectroscopy maps the folding landscape
  of a large protein.
\newblock \emph{Nature Commun.}, 2:\penalty0 493, 2011.

\bibitem[Keller et~al.(2014)Keller, Kobitski, J{\"a}schke, Nienhaus, and
  No{\'e}]{KellerNoe_JACS14_HMM-FRET}
B.~G. Keller, A.~Y. Kobitski, A.~J{\"a}schke, U.~G. Nienhaus, and F.~No{\'e}.
\newblock Complex rna folding kinetics revealed by single molecule fret and
  hidden markov models.
\newblock \emph{J. Am. Chem. Soc.}, 136:\penalty0 4534--4543, 2014.

\bibitem[Gopich et~al.(2009)Gopich, Nettels, Schuler, and
  Szabo]{GopichSzabo_FRETCorrelation_JCP09}
I.~V. Gopich, D.~Nettels, B.~Schuler, and A.~Szabo.
\newblock {Protein dynamics from single-molecule fluorescence intensity
  correlation functions}.
\newblock \emph{J. Chem. Phys.}, 131\penalty0 (9):\penalty0 095102, 2009.

\bibitem[Gopich and Szabo(2009)]{GopichSzabe_HMMFRET_JCP09}
I.~V. Gopich and A.~Szabo.
\newblock {Decoding the pattern of photon colors in single-molecule FRET.}
\newblock \emph{J. Phys. Chem. B}, 113\penalty0 (31):\penalty0 10965--10973,
  2009.

\bibitem[Jazani et~al.(2019)Jazani, Sgouralis, and
  Press{\'e}]{JazaniEtAl_JCPMSM19_Tracking}
S.~Jazani, I~Sgouralis, and S.~Press{\'e}.
\newblock A method for single molecule tracking using a conventional
  single-focus confocal setup.
\newblock \emph{J. Chem. Phys.}, 150:\penalty0 125101, 2019.

\bibitem[Ribeiro et~al.(2018)Ribeiro, Bravo, Wang, and
  Tiwary]{RibeiroTiwary_JCP18_RAVE}
J.~M.~L. Ribeiro, P.~Bravo, Y.~Wang, and P.~Tiwary.
\newblock Reweighted autoencoded variational bayes for enhanced sampling
  (rave).
\newblock \emph{J. Chem. Phys.}, 149:\penalty0 072301, 2018.

\bibitem[Zhang et~al.(2019)Zhang, Yang, and
  No{\'e}]{ZhangYanNoe_ChemRxiv19_TALOS}
J.~Zhang, Y.~I. Yang, and F.~No{\'e}.
\newblock Targeted adversarial learning optimized sampling.
\newblock \emph{ChemRxiv. DOI: 10.26434/chemrxiv.7932371}, 2019.

\bibitem[Hern{\'a}ndez et~al.(2018)Hern{\'a}ndez, Wayment-Steele, Sultan,
  Husic, and Pande]{Hernandez_PRE18_VTE}
C.~X. Hern{\'a}ndez, H.~K. Wayment-Steele, M.~M. Sultan, B.~E. Husic, and B.~S.
  Pande.
\newblock Variational encoding of complex dynamics.
\newblock \emph{Phys. Rev. E}, 97:\penalty0 062412, 2018.

\bibitem[Jung et~al.(2019)Jung, Covino, and Hummer]{Jung2019}
H.~Jung, R.~Covino, and G.~Hummer.
\newblock Artificial intelligence assists discovery of reaction coordinates and
  mechanisms from molecular dynamics simulations.
\newblock \emph{arXiv:1901.04595}, 2019.

\bibitem[Wang et~al.(2019)Wang, Ribeiro, and
  Tiwary]{WangEtAl_NatComm19_PastFutureBottleneck}
Y.~Wang, J.~M.~L. Ribeiro, and P.~Tiwary.
\newblock Past--future information bottleneck for sampling molecular reaction
  coordinate simultaneously with thermodynamics and kinetics.
\newblock \emph{Nat. Commun.}, 10:\penalty0 3573, 2019.

\bibitem[Harmeling et~al.(2003)Harmeling, Ziehe, Kawanabe, and
  M\"{u}ller]{HarmelingEtAl_NeurComput03_KernelTDSEP}
S.~Harmeling, A.~Ziehe, M.~Kawanabe, and K.-R. M\"{u}ller.
\newblock Kernel-based nonlinear blind source separation.
\newblock \emph{Neur. Comp.}, 15:\penalty0 1089--1124, 2003.

\bibitem[Schwantes and Pande(2015)]{SchwantesPande_JCTC15_kTICA}
C.~R. Schwantes and V.~S. Pande.
\newblock Modeling molecular kinetics with tica and the kernel trick.
\newblock \emph{J. Chem. Theory Comput.}, 11:\penalty0 600--608, 2015.

\bibitem[Coifman et~al.(2005)Coifman, Lafon, Lee, Maggioni, Nadler, Warner, and
  Zucker]{CoifmanLafon_PNAS05_DiffusionMaps}
R.~R. Coifman, S.~Lafon, A.~B. Lee, M.~Maggioni, B.~Nadler, F.~Warner, and
  S.~W. Zucker.
\newblock Geometric diffusions as a tool for harmonic analysis and structure
  definition of data: Diffusion maps.
\newblock \emph{Proc. Natl. Acad. Sci. USA}, 102:\penalty0 7426--7431, 2005.

\bibitem[Rohrdanz et~al.(2011)Rohrdanz, Zheng, Maggioni, and
  Clementi]{RohrdanzClementi_JCP134_DiffMaps}
M.~A. Rohrdanz, W.~Zheng, M.~Maggioni, and C.~Clementi.
\newblock Determination of reaction coordinates via locally scaled diffusion
  map.
\newblock \emph{J. Chem. Phys.}, 134:\penalty0 124116, 2011.

\bibitem[Klus et~al.(2019)Klus, Bittracher, Schuster, and
  Sch{\"u}tte]{KlusEtAl_JCPMSM19_Kernel}
S.~Klus, A.~Bittracher, I.~Schuster, and C.~Sch{\"u}tte.
\newblock A kernel-based approach to molecular conformation analysis.
\newblock \emph{J. Chem. Phys.}, 150:\penalty0 244109, 2019.

\bibitem[{de Sancho} and Aguirre(2019)]{DeSancho_JCIM19_MasterMSM}
D.~{de Sancho} and A.~Aguirre.
\newblock Mastermsm: A package for constructing master equation models of
  molecular dynamics.
\newblock \emph{J. Chem. Inf. Model.}, 59:\penalty0 3625--3629, 2019.

\bibitem[Porter et~al.(2019)Porter, Zimmerman, and
  Bowman]{PorterEtAl_JCPMSM19_Enspara}
J.~R. Porter, M.~I. Zimmerman, and G.~R. Bowman.
\newblock Enspara: Modeling molecular ensembles with scalable data structures
  and parallel computing.
\newblock \emph{J. Chem. Phys.}, 150:\penalty0 044108, 2019.

\bibitem[Schulz et~al.(2019)Schulz, von Hansen, Daldrop, Kappler, No{\'e}, and
  Netz]{SchulzEtAl_JCPMSM19_DNA}
R.~Schulz, Y.~von Hansen, J.~O. Daldrop, J.~Kappler, F.~No{\'e}, and R.~R.
  Netz.
\newblock Collective hydrogen-bond rearrangement dynamics in liquid water.
\newblock \emph{J. Chem. Phys.}, 150:\penalty0 244504, 2019.

\bibitem[Gopich and Szabo(2019)]{GopichSzabo_JCPMSM19_TwoSite}
I.~V. Gopich and A.~Szabo.
\newblock Diffusion-induced competitive two-site binding.
\newblock \emph{J. Chem. Phys.}, 150:\penalty0 094104, 2019.

\bibitem[Shin and Kolomeisky(2019)]{ShinKolomeisky_JCPMSM19_Walker}
J.~Shin and A.~B. Kolomeisky.
\newblock Molecular search with conformational change: One- dimensional
  discrete-state stochastic model.
\newblock \emph{J. Chem. Phys.}, 150:\penalty0 174104, 2019.

\bibitem[Pinamonti et~al.(2019)Pinamonti, Paul, No{\'e}, Rodriguez, and
  Bussi]{PinamontiEtAl_JCPMSM19_DNA}
G.~Pinamonti, F.~Paul, F.~No{\'e}, A.~Rodriguez, and G.~Bussi.
\newblock The mechanism of rna base fraying: Molecular dynamics simulations
  analyzed with core-set markov state models.
\newblock \emph{J. Chem. Phys.}, 150:\penalty0 154123, 2019.

\bibitem[Chakraborty and Wales(2019)]{ChakrabortyWales_JCPMSM19_DPS}
D.~Chakraborty and D.~J. Wales.
\newblock Dynamics of an adenine- adenine rna conformational switch from
  discrete path sampling.
\newblock \emph{J. Chem. Phys.}, 150:\penalty0 125101, 2019.

\bibitem[Wales(2002)]{Wales_MolPhys100_3285}
D.~J. Wales.
\newblock {Discrete path sampling}.
\newblock \emph{Mol. Phys.}, 100:\penalty0 3285--3305, 2002.

\bibitem[Sengupta et~al.(2019)Sengupta, Carballo-Pacheco, and
  Strodel]{SenguptaEtAl_JCPMSM19_MFPT}
U.~Sengupta, M.~Carballo-Pacheco, and B.~Strodel.
\newblock Automated markov state models for molecular dynamics simulations of
  aggregation and self- assembly.
\newblock \emph{J. Chem. Phys.}, 150:\penalty0 115101, 2019.

\bibitem[Ziehe et~al.(2004)Ziehe, Laskov, Nolte, and
  M\"{u}ller]{ZieheEtAl_JMLR04_FastTDSEP}
A.~Ziehe, P.~Laskov, G.~Nolte, and K.-R. M\"{u}ller.
\newblock A fast algorithm for joint diagonalization with non-orthogonal
  transformations and its application to blind source separation.
\newblock \emph{J. Mach. Learn. Res.}, 5:\penalty0 777--800, 2004.

\end{thebibliography}

\end{document}